\documentclass[12pt,a4paper]{article}%
\usepackage{amsmath}
\usepackage{amsfonts}
\usepackage{amssymb}
\usepackage{graphicx}%
\providecommand{\U}[1]{\protect\rule{.1in}{.1in}}

\begin{document}

\title{CUSUM Average Run Length: Conditional or Unconditional?}
\author{Fred Lombard\\Department of Statistics\\University of Johannesburg\\fredl@uj.ac.za
\and Douglas M. Hawkins\\University of Minnesota\\Minneapolis, MN\\dhawkins@umn.edu }
\date{}
\maketitle

\begin{abstract}
The behavior of CUSUM charts depends strongly on how they are initialized.
Recent work has suggested that self-starting CUSUM methods retain some
dependence on their very first readings, and introduced the concept of
"conditional average run length" (CARL) -- the average run length conditioned
on the first few process readings -- as a result of which is it claimed that
different practitioners using the same methodology could experience different
ARLs because of the random differences in their earliest readings. We cast
doubt on whether CARL is relevant to practitioners who use self-starting
methods and argue that the unconditional ARL is the relevant measure there.

\end{abstract}

\section{Introduction}

Cumulative sum (CUSUM) charts are highly sensitive to even small shifts in the
process parameters, and their behavior therefore depends strongly on how they
are initialized. The traditional approach of plugging in estimates from a
Phase I is well known to require inordinately large and unproductive Phase I
samples (Jones, Champ and Rigdon, 2004, Jensen, et al., 2006), and this has
led to the development of "self-starting" approaches which circumvent the need
for large Phase I samples but nevertheless control average run length
behavior. In particular, the in-control average run length (ARL) of
self-starting methods is exactly that obtained in the known-parameter setting.
Recent work (Keefe, Woodall and Jones-Farmer, 2015) has highlighted the
concept of "conditional average run length" (CARL) -- the average run length
conditioned on the first few process readings -- and practitioner to
practitioner variability, motivated by the idea that different practitioners
using the same self-starting methodology could experience different CARLs
because of the random differences in their earliest readings.

The exemplar cumulative sum control chart (CUSUM) is that defined by Page
(1954) and explored in the monograph by Hawkins and Olwell (1998). It deals
with a sequence of independent $N(\mu,\sigma^{2})$ readings $X_{1}%
,X_{2},...X_{m}...$. In its standardized form rescaling the data to N(0,1),
the CUSUM defines
\begin{align*}
U_{n}  &  =(X_{n}-\mu)/\sigma\\
D_{0}  &  =0
\end{align*}
and $D_{0}=0$,
\[
D_{n}=max(0,D_{n-1}+U_{n}-\delta)
\]
for $n\geq1$, where $\delta$ denotes the reference value. This canonical form
requires that $\mu$ and $\sigma$ be known. The traditional resolution to this
was to conduct a separate preliminary Phase I study, and use its mean
$\hat{\mu}$ and standard deviation $\hat{\sigma}$ in place of $\mu$ and
$\sigma$, replacing $U_{m}$ in the CUSUM defining equation by $W_{m}%
=(X_{m}-\hat{\mu})/\hat{\sigma}.$ However $W_{m}$ does not follow the N(0,1)
distribution, nor are the successive $W_{m}$ independent thanks to their
common dependence on $\hat{\mu}$ and $\hat{\sigma}$. Consequently, when the
control limits applicable to the known parameter case are used, the run
lengths of this plug-in CUSUM differ substantially from those of the
known-parameter setting unless many hundreds of readings are included in the
Phase I study.

Hawkins (1987) and Quesenberry (1991) - see also Zantek (2006) - proposed the
self-starting methodology in which the estimates of $\mu$ and $\sigma$ are
updated with each new observation. Write, for $n\geq2$,%

\[
\hat{\mu}_{n}=\frac{X_{1}+\cdots+X_{n}}{n},\ \ \hat{\sigma}_{n}^{2}%
=\frac{(X_{1}-\hat{\mu}_{n})^{2}+\cdots+(X_{m}-\hat{\mu}_{n})^{2}}{n-1}%
\]
and for $n\geq m\geq3$, $W_{n}=(X_{n}-\hat{\mu}_{n-1})/\hat{\sigma}_{n-1}$.
CUSUMs can be constructed starting from $m\geq3$. These $m$ initial
observations constitute the \textquotedblleft warmup" of a self-starting
scheme, corresponding to the traditional Phase I study, from which it segues
smoothly into the Phase II online monitoring by updating the estimates of
$\mu$ and $\sigma$ upon the arrival of each new observation. The random
variable $W_{n}$ follows a scaled $t$ distribution with $n-2$ degrees of
freedom and the successive $W_{n},\ n\geq m$ are statistically independent.
They may be transformed to independent exact $N(0,1)$ quantities $U_{n}$ by
the double probability integral transform%
\begin{equation}
U_{n}=\Phi^{-1}(F_{n-2}(W_{n}\sqrt{(n-1)/n}) \label{self-starting}%
\end{equation}
where $\Phi^{-1}$ is the inverse normal CDF and $F_{n-2}$ is the cumulative
distribution function of Student's $t$ with $n-2$ degrees of freedom. The
CUSUM\ recursion is $D_{0}=0$ and%

\[
D_{k}=max(0,D_{k-1}+U_{m+k}-\delta)
\]
for $k\geq1$. The run length, $N$, is the first $k\geq1$ at which $D_{k}$
exceeds the control limit. A crucial requirement underlying the self-starting
method is that fresh warmup observations be taken upon every restart of the
procedure. The effect of this requirement is that the successive
$U_{n},\ n\geq m$ are statistically independent with standard normal
distributions and that the known-parameter CUSUM's control limits can be used.
The in-control ARL of the self-starting CUSUM will then equal that of the
known-parameter CUSUM.

\section{Practitioner-to-practitioner variability}

Keefe, et al. (2015) argue that practitioners who use self-starting CUSUMS
will experience different Phase II in-control ARLs simply because their warmup
estimates of $\mu$ and $\sigma$ differ and (page 496) that this points to a
possible defect in the self-starting CUSUM,\textbf{ }to wit: "It is not the
case, as stated by Hawkins and Olwell (1998, p. 162) that the self-starting
approach removes the estimation issue from the problem
completely.\textquotedblright.

The key issue in understanding the relevance of the CARL and
practitioner-to-practitioner variation concepts is in how the CUSUM is
re-initialized following a signal. We identify two scenarios:

\begin{description}
\item[Scenario 1:] In each run gather\textit{ }$m$\textit{ }fresh warmup
readings and restart the CUSUM.
\end{description}

\begin{description}
\item[Scenario 2:] Keep the original $m$ warmup readings and their resulting
$\hat{\mu}_{m}$ and $\hat{\sigma}_{m}$ and restart every run from that baseline.
\end{description}

In what follows it is assumed that the underlying process is in control. The
run length of the self-starting CUSUM is denoted by $N$ and the warmup data by
$X_{m}$\emph{.}The self-starting method demands a Scenario 1 initialization of
every run, that is, $X_{m}$ varies from run to run. Then the run lengths
$N_{1},\ N_{2},\ \ldots$observed in a long series of runs are i.i.d. copies of
$N$ and their average will converge to the nominal in-control ARL $E[N]$.

Keefe, et Al. (2015), consider the conditional ARL $E\left[  N|\mathbf{X}%
_{m}\right]  $ and find in a Monte Carlo study that it varies substantially
with $X_{m}$, as would be expected because $E\left[  N|\mathbf{X}_{m}\right]
$ is a random variable that depends upon $X_{m}$. Different practitioners will
therefore have different warmup sets $X_{m}$ and the ARL observed by a
practitioner whose warmup set is $X_{m}=x_{m}$ is $E\left[  N|\mathbf{X}%
_{m}=x_{m}\right]  $. But this is the ARL of the self-starting CUSUM run under
scenario 2 initialization. With this initialization the $U_{i}$ are not
statistically independent, nor do they have standard normal distributions.
Thus, none of the CUSUMs being run by the various practitioners is the
self-starting CUSUM defined by Hawkins (1987). It is therefore not at all
clear how the behaviour of these CUSUMs can be construed as indicating a
defect in the Hawkins (1987) self-starting CUSUM.

The link between the unconditional ARL of the self-starting CUSUM and its
CARLs considered by Keefe, et Al. (2015) is provided by the well known
formula,%
\begin{equation}
E\left[  N\right]  =E\left[  E\left[  N|\mathbf{X}_{m}\right]  \right]
,\label{Tower law}%
\end{equation}
that is, the unconditional ARL $E\left[  N\right]  $ is equal to the long run
average of the CARLs $E\left[  N|\mathbf{X}_{w}\right]  $ over all warmup sets
$X_{m}$. The warmup readings $\mathbf{X}_{m}$ define the CARL of the CUSUM
but, as seen from (\ref{Tower law}), this CARL is a notional rather than a
real construct. The process owner who implements the self-starting CUSUM
correctly, i.e. with Scenario 1 initializations, will see a single run length
from any particular warmup set $\mathbf{X}_{m}$ after which that warmup set is
discarded and a new one generated. In accordance with (\ref{Tower law}), the
average of the process owner's observed run lengths will therefore be the
unconditional ARL, which equals the nominal value, no matter how the
individual CARLs $E\left[  N|\mathbf{X}_{w}\right]  $ behave. The conclusion,
referred to above, reached by Keefe, et Al. (2015) is therefore unwarranted.

In fact, the practitioner to practitioner concept can be used to cast doubt on
the validity of many perfectly valid statistical procedures. As a case in
point, consider the two-sample t-test for the equality of the means in two
populations with a common but unknown variance $\sigma^{2}$. The test
statistic is%
\[
T=\frac{|\bar{X}-\bar{Y}|}{S}%
\]
where $\bar{X}_{m}$ and $\bar{Y}_{n}$ denote the sample means and
\[
S^{2}=\frac{%
{\textstyle\sum\nolimits_{i=1}^{m}}
(X_{i}-\bar{X}_{m})^{2}+%
{\textstyle\sum\nolimits_{i=1}^{m}}
(Y_{i}-\bar{Y}_{n})^{2}}{m+n-2}%
\]
denotes the pooled estimate of $\sigma^{2}$. The null hypothesis is rejected
at the $100\alpha\%$ level of significance whenever $T$ exceeds the upper
$100(1-2\alpha)\%$ point of the Student t distribution with $m+n-2$ degrees of
freedom. However, conditional upon $\bar{Y}=y$, the level of significance is
no longer $100\alpha\%$ but something quite different and it will vary with
the $y$ values observed by different practitioners. Clearly, this fact cannot
serve as justification for a statement that "It is not the case that the two
sample t test removes the variance estimation issue from the problem
completely\textquotedblright. The t-test requires that fresh $X$ and $Y$
samples be used in every repetition of the test (scenario 1) while the
conditional test restricts the mean of the $Y$ sample to a fixed value in
every repetition of the test (scenario 2).

We conclude that the CARL and practitioner-to-practitioner variation concepts
are not relevant to anyone using the self-starting CUSUM of Hawkins (1987) and
that this CUSUM, based upon (\ref{self-starting}) does indeed solve the
estimation problem completely.

A fully satisfactory parametric self-starting CUSUM is presently available
only when the underlying distribution is normal. In other multi-parameter
distributions the self-starting property applies only to a single parameter -
the remaining parameters must be known. If these are estimated from Phase I
data, the self-starting property is lost and we again have a plug-in CUSUM.
Thus, it is perhaps appropriate to point out that there are available
attractive distribution-free alternatives to some plugin CUSUMs. By
distribution-free is meant that the in-control properties of the CUSUM do not
depend upon a parametric specification of the underlying distribution. For
instance, a straightforward approach to the problem is to use sequential rank
CUSUMs (Lombard and Van Zyl, 2018 and Van Zyl and Lombard, 2018). Besides
being distribution-free and not requiring any parameter estimates, hence no
warmup data, methods are available for estimating, a priori, the
out-of-control ARL of these CUSUMs. Furthermore, "once and for all situations"
control limits are available. Since no parameter estimates or warmup data are
required, the concept of practitioner-to-practitioner variability is vacuous
in these CUSUMs. The Wilcoxon-type CUSUM of Hawkins and Deng (2010) does not
require any parameter estimates and is completely distribution-free. Since no
parameter estimates are required, the concept of practitioner-to-practitioner
variability is vacuous in these CUSUMs.

Again in nonparametric settings, Chatterjee and Qiu (2009), Gandy and Kvaloy
(2013) and Saleh, et al. (2016) show how the bootstrap may be used to obtain
control limits that would yield ARLs close to the nominal value when a
substantial amount of Phase I data are available. However, since the control
limits depend on the (unknown) underlying distribution, "once and for all
situations" control limits do not exist. Furthermore, in any given
application, new control limits must be generated whenever the in-control mean
or variance has undergone a permanent shift. The normal self-starting and
distribution-free CUSUMs (\ref{self-starting}) are not affected by such shifts
and use "once and for all situations" control limits.

Some useful further work would entail comparisons between the out-of control
behaviors of the rank-based CUSUMs and the bootstrap-defined plug-in CUSUMS.


\begin{thebibliography}{99}                                                                                               %


\bibitem {Chatterjee}Chatterjee, S. and Qiu, P. (2009).Distribution-Free
Cumulative Sum Control Charts Using Bootstrap-based Control Limits.
\textit{The Annals of Applied Statistics}, \textbf{3}, 349-369.

\bibitem {Gandy}Gandy, A. and Kvaloy, J.T. (2013). Guaranteed Conditional
Performance of Control Charts via Bootstrap Methods. \textit{Scandinavian
Journal of Statistics}, \textbf{40}, 647-668.

\bibitem {Hawkins}Hawkins, D.M. (1987). Self-Starting CUSUM Charts for
Location and Scale. \textit{Journal of the Royal Statistical Society. Series
D}, \textbf{36}, 299-316.

\bibitem {Jensen}Jensen, W.A., Jones-Farmer, L.A., Champ, C.W. and Woodall,
W.H., (2006). Effects of parameter estimation on control chart properties: A
literature review. \textit{Journal of Quality Technology}, \textbf{38} (4), 349-364.

\bibitem {Jones}Jones, A.L., Champ, C.W. and S.E. Rigdon, (2004). The Run
Length Distribution of the CUSUM with Estimated Parameters. \textit{Journal of
Quality Technology}, \textbf{36} (1), 95-108.

\bibitem {Keefe et al}Keefe, M.J., Woodall, W.H. and Jones-Farmer, L.A.
(2015). The Conditional In-Control Performance of Self-Starting Control
Charts. \textit{Quality Engineering}, \textbf{27}, 488-499.

\bibitem {HO}Hawkins, D.M. and Olwell, D.H. (1998). \textit{Cumulative Sum
Charts and Charting for Quality Improvement}, Springer New York.

\bibitem {HD}Hawkins, D.M. and Deng, Q. (2010). A Nonparametric Change-Point
Control Chart. \textit{Journal of Quality Technology}, \textbf{42}, 165-173.

\bibitem {L and VZ}Lombard, F. and Van Zyl, C. (2018). Signed Sequential Rank
CUSUMs. \textit{Computational Statistics and Data Analysis}, \textbf{118}, 30-39.

\bibitem {Quesenberry}Quesenberry, C.P., (1991). SPC Q charts for start-up
processes and short or long runs. \textit{Journal of Quality Technology},
\textbf{23} (3), 213-224.

\bibitem {Saleh et al}Saleh, N.A., Zwetsloot, I.M., Mahmoud, A.M. and Woodall,
W.H. (2016). CUSUM charts with controlled conditional performance under
estimated parameters. \textit{Quality Engineering}, \textbf{28}, 402-425.

\bibitem {VZ and L}Van Zyl. C. and Lombard, F., (2018). Sequential Rank CUSUMs
for Location and Dispersion. \textit{South African Statistical Journal,}
\textbf{52}, 93-113.

\bibitem {Zantek}Zantek, F. (2006). Design of Cumulative Sum Schemes for
Start-Up Processes and Short Runs. \textit{Journal of Quality Technology},
\textbf{38} (4), 365-375.
\end{thebibliography}
\end{document}